**Experimental demonstration of fiber-accessible metal nanoparticle plasmon waveguides for planar energy guiding and sensing**

Stefan A. Maier*, Michelle D. Friedman, Paul E. Barclay, and Oskar J. Painter
*Thomas J. Watson Laboratory of Applied Physics, California Institute of Technology*
*Pasadena, CA 91125, USA*

**Abstract**

Experimental evidence of mode-selective evanescent power coupling at telecommunication frequencies with efficiencies up to 75 % from a tapered optical fiber to a carefully designed metal nanoparticle plasmon waveguide is presented. The waveguide consists of a two-dimensional square lattice of lithographically defined Au nanoparticles on an optically thin silicon membrane. The dispersion and attenuation properties of the waveguide are analyzed using the fiber taper. The high efficiency of power transfer into these waveguides solves the coupling problem between conventional optics and plasmonic devices and could lead to the development of highly efficient plasmonic sensors and optical switches.

Corresponding Author:

Stefan A. Maier
California Institute of Technology
Mail-Code 128-95
Pasadena, CA 91125, USA
Tel 626 395 8008
Fax 626 795 7258
e-mail: stmaier@caltech.edu

* electronic address: stmaier@caltech.edu

Micro- and nanostructured assemblies of metals sustaining coherent, non-radiative electronic excitations known as surface plasmon polaritons [1] are an intriguing materials system promising a wide range of applications in photonics and telecommunications [2]. Basic building blocks of plasmonic devices will be planar metallic waveguides such as micron-sized stripes [3], nanowires [4] and nanoparticle arrays [5] for the transport of electromagnetic energy with a mode confinement on the wavelength and subwavelength scale, respectively. However, the excitation of spatially localized plasmons in such waveguides poses a considerable challenge due to the small size and complex mode-shape of the guided modes, resulting in a huge spatial mismatch with diffraction-limited optical beams and conventional waveguides. For far-field excitation of localized plasmons in thin Au films, coupling efficiencies of about 15% have been estimated [6], while the excitation of subwavelength-scale modes in one-dimensional nanoparticle waveguides typically employs near-field fiber probes with light throughputs below 0.1% [7]. Moreover, neither of the two methods is mode-selective. Using a novel design concept for metal nanoparticle plasmon waveguides based on silicon-on-insulator technology, we present experimental observations of mode-selective energy transfer at telecommunication frequencies from a conventional fiber taper to a plasmon waveguide with efficiencies up to 75 %. This concept is extendable to higher frequencies and promises applications in energy guiding and optical sensing with high efficiencies.

The demonstration of non-diffraction-limited guiding of electromagnetic energy over micron- and sub-micron distances in resonantly excited metallic nanowires [4] and nanoparticle waveguides [5] has recently been achieved as a first step towards the ultimate goal of building highly integrated photonic circuits for channeling energy to nanoscale detectors. Due to the heightened local fields surrounding metallic guiding structures, such devices could find useful applications not only in photonics and telecommunications but also in biological sensing [8] of molecules in localized "hot-spots". For this vision to come true, new ways of coupling light into such devices have to be developed that are more efficient than non-mode selective techniques such as diffraction-limited far-field coupling and local excitation using nanoscopic probes. On a larger scale, the non-resonant guiding of electromagnetic energy at visible [9] and near-infrared frequencies [3, 10] using micron-scale metallic stripes has been demonstrated over distances up to several hundred microns employing far-field and end-fire excitation techniques, respectively. However, such long-ranging surface plasmon polaritons can only be sustained and efficiently excited for stripes embedded in a symmetric environment [11], making them of limited use for sensing applications and also complicating the employment of these structures as input ports to sub-diffraction-limit plasmon waveguides.

To overcome this challenging obstacle, we have developed a new design concept of a metal nanoparticle plasmon waveguide that can be efficiently excited via evanescent, phase-matched coupling using small tapers drawn from standard silica fibers [12, 13]. The waveguide consists of a hybrid structure of SOI (silicon-on-insulator) and a lithographically defined square lattice of metal nanoparticles on the optically thin, undercut silicon membrane. In order to allow for non-resonant excitation of the metal nanoparticles to reduce the absorptive heating losses without a concomitant increase in radiative loss, we employ a lateral grading in nanoparticle size to confine the mode to the centre of the waveguide (Note that for one-dimensional metal nanoparticle chains, resonant excitation with significantly higher heating losses is necessary [14] in order to tightly confine the mode to the sub-wavelength guiding structure). Vertically, the confinement is ensured both by bound metal/air surface plasmons and the undercut geometry of the silicon membrane. Figure 1a shows an example of a finite-difference time-domain calculated optical mode profile (electric field intensity) for the fundamental waveguide mode in top and side geometry. Further details on the design will be reported elsewhere [15].

Due to the periodicity of the structure in the propagation direction, the dispersion relation of the plasmon polariton modes is folded back into the first Brillouin zone. This way, the upper plasmon band crosses both the light-lines of silica and air, suggesting the possibility of phase-matched, mode-selective excitations of the waveguide using silica fibre tapers [13] placed in close proximity to the waveguide. In order to spatially and spectrally investigate energy transfer in this system, we fabricated a plasmon waveguide designed to allow efficient evanescent coupling with a fiber taper at a wavelength of 1.6 µm in the telecommunication band. Figure 1b shows scanning electron micrographs of the fabricated waveguide based upon a 530 nm square lattice of Au nanoparticles on a 250 nm Si membrane created by undercutting the 2 µm silica layer of a SOI wafer using hydrofluoric acid. The nanoparticles are 50 nm in height, and have a diameter of approximately 250 nm in the centre of the waveguide with a linear lateral grading down to 210 nm after three of the six lateral periods.

A fiber taper of approximately 1.5 µm diameter was fabricated by simultaneously heating and stretching a segment of a standard single mode telecommunication fiber. Near-infrared light of a scanning external cavity laser with a wavelength range of 1565 – 1625 nm was coupled into the fiber, and the transmitted light detected with a photodetector. The thinnest part of the fiber taper was positioned above and parallel to the plasmon waveguide. At a taper height of several microns above the centre of the waveguide, the taper mode did not interact with the waveguide, resulting in unity transmission over the entire wavelength range. The taper was then lowered until a resonant drop in transmission indicative of coupling to the plasmon waveguide was observed. Lateral movement of the taper at constant height allowed then the examination of the spatial coupling profile. Figure 2a shows a contour plot of the power transmitted through the fiber taper on a linear color scale versus wavelength and lateral taper position. Several spectral and spatial features are clearly discernible: The drop in transmitted power centered around 1590 nm when the taper is placed over the central part of the waveguide is indicative of coupling to the fundamental plasmon waveguide mode. Upon taper displacement away from the waveguide centre, the 1590 nm fundamental mode is seen to disappear, and a periodic modulation of the transmitted signal occurs (Figure 2, dotted vertical lines A and B). We attribute this modulation to enhanced non-resonant coupling into the underlying Si slab mediated by the metal nanoparticle lattice. Also, a second, weaker transmission dip around 1570 nm is discernable in the transmission data when the taper is placed over the outer parts of the waveguide. We believe these dips to be due to resonant excitation of the higher order transverse odd waveguide mode, with the different depths of the transmission dips on either side of the waveguide probably being due to small height variations during lateral movement of the taper. Note that both for the fundamental and the higher order transverse mode resonances, the periodic modulations of the transmitted signal due to taper-slab coupling are suppressed. When the taper is shifted completely off the plasmon waveguide and sits entirely over the Si slab, the transmission almost completely recovers to unity, but for coupling into the Si membrane as evidenced by modulations of the same periodicity but weaker amplitude as for the metal-assisted scattering. Figure 2b summarizes these results by showing the lateral transmission profile at the centre wavelength of the fundamental mode (1590 nm, upper graph) and that of the next higher order mode (1570 nm, lower graph). We have thus demonstrated that fiber tapers allow for mode-selective coupling to a carefully designed plasmon waveguide, giving this method significant advantage over other excitation geometries such as near-field probes and far-field coupling. In our experiment, the spatial full-width at half-maximum of approximately 4 μm of the fundamental mode was rather large, due to convolution with the finite-width of the fiber taper and contributions from the metal-assisted scattering discussed above.

The dispersion of the fundamental mode was further examined by translating the fiber in the direction of the waveguide in steps of 200 μm, thereby varying the fiber diameter and thus the wave vector of phase-matching. Figure 3a shows the spectral position of the fundamental resonance (white dots correspond to minimum in transmission depth) overlaid on a contour plot of the transmitted signal versus taper position along the waveguide direction and wavelength. In order to adjust for varying gaps between the taper and the waveguide with changing taper diameter, the transmission data have been normalized to zero transmission in the centre of the respective resonances. The red-shift of the resonance with increasing fiber diameter as depicted in the inset is indicative of contra-directional coupling. Thus phase-matching does indeed take place between the zone-folded upper plasmon band and the taper light line.

The efficiency of power transfer into our plasmon waveguide was studied using a newly drawn fiber taper of 1.1 µm diameter. Figure 3b shows a representative wavelength scan of the transmitted signal at a height of approximately 1100 nm above the waveguide centre after extensive optimization of the alignment of the taper with respect to the waveguide. For this height, the normalized transmitted signal dropped to 0.12 at the centre of the resonance dip, while recovering back to 0.75 at the higher frequency end of the laser scan range (red curve). Using a second laser with a scan range between 1510 and 1580 nm, a nearly complete recovery of the transmission to about 0.87 was confirmed (blue curve). We have thus observed a power transfer of about 75 % from the fibre taper to the plasmon waveguide [16].

In all our experiments, the depth of the resonance dip in transmission was seen to monotonically increase with decreasing taper height as expected for contra-directional coupling (i.e. no Rabi-type flopping of power as in co-directional coupling). The inset of Figure 3b shows examples of wavelength scans of the transmitted power for different taper heights, decreasing from 4 µm to zero gap (defined as the gap where taper is touching the surface) in 500 nm steps. Note that the wavelength dependence of the coupling is broader and not as sharply defined as in an ideal contra-directional coupler, which we believe to be due to fabrication non-idealities.

The occurrence of Fabry-Perot resonances due to reflections at the mirror ends of our waveguide superimposed on the transmission data allowed the estimation of the group velocity and energy decay length of the propagating plasmon mode. By analyzing the spacing and width of the resonances, the group velocity was thus estimated to about 15 % of the velocity of light in vacuum (corresponding to a group index of 6.67) in good agreement with finite-difference time-domain simulations [15], and the 1/e energy decay length to about 50 μm. Note that the decay length estimate assumes perfect mirror reflectivities. Due to almost certain reflectivities less than unity, the decay length of 50 μm is thus a lower estimate only.

At this point, it is worth noting that our design concept can also be scaled to higher frequencies towards the visible regime of the spectrum by an appropriate change in lattice constant. The higher absorptive losses for near-resonant excitations at lower wavelengths can then be partially counteracted by a change of the materials system to silver. The high efficiency of power transfer into our plasmon waveguide should thus allow intriguing applications at visible and near-infrared frequencies. For example, the use as a coupling structure to other planar plasmonic devices such as cavities and resonantly excited one-dimensional particle waveguides as presented in [5] can be envisioned. Moreover, the good accessibility of the optical surface mode suggests high promise for this design as an efficient optical sensor for biological agents.

In summary, we have presented a novel plasmon waveguide based on a two-dimensional lattice of Au nanoparticles on a thin silicon membrane fabricated using silicon-on-insulator processing techniques that can be efficiently excited using fiber tapers. Mode-selective power transfer efficiencies up to 75 % have been demonstrated. Such waveguides could make possible the interconnection of a wealth of recently proposed plasmonic micro- and nanostructures with conventional fiber optics and also lead to the design of high-performance plasmonic sensors.

Acknowledgements

The authors are grateful to M. Borselli and T. J. Johnson for help with fabrication.

Figure Captions:

Figure 1 (color). a) Electric field intensity distribution of a plasmon waveguide mode in top (upper plot) and side view (lower plot) calculated using finite-difference time-domain simulations. b) Scanning electron micrograph of a fabricated plasmon waveguide on an undercut Si membrane. The waveguide is terminated by mirrors at both ends.

Figure 2 (color). a) Contour plot showing the transmitted power through the fiber taper on a linear color scale versus wavelength and lateral position of the taper with respect to the waveguide. b) Lateral transmission profile at a wavelength of 1590 nm (fundamental mode, upper graph) and 1570 nm (higher order mode, lower graph).

Figure 3 (color). a) Contour plot showing the normalized transmitted power through the fiber taper on a linear color scale versus wavelength and taper displacement in the direction of the waveguide. The transmission minimum (marked by white dots) is seen to move towards larger wavelength for thicker fiber diameters as depicted in the inset, indicative of contra-directional coupling from the fiber taper to the plasmon waveguide. b) Wavelength scan of the transmitted power through the fiber taper placed in close proximity over the centre of the plasmon waveguide obtained using two different external cavity lasers with overlapping scan ranges. The inset shows the evolution of the coupling when the taper is successively lowered towards the waveguide in 50 nm steps (every tenth step shown).

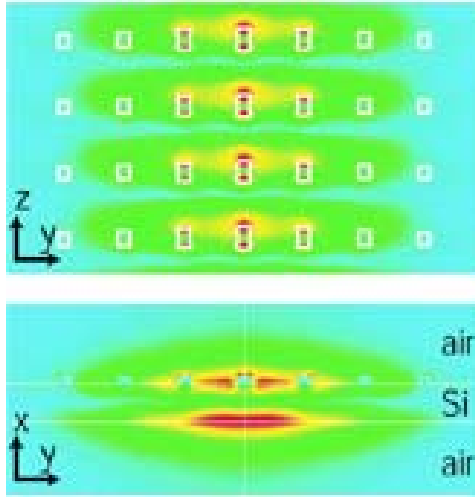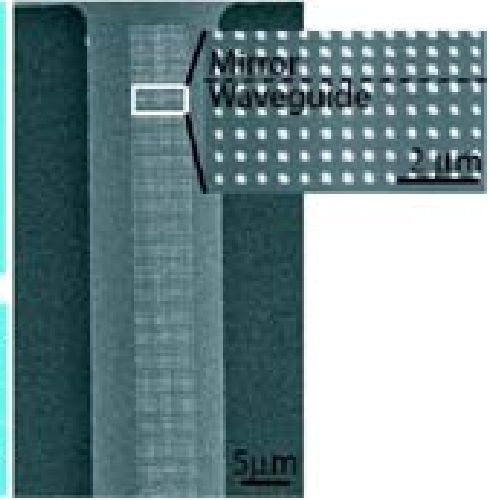

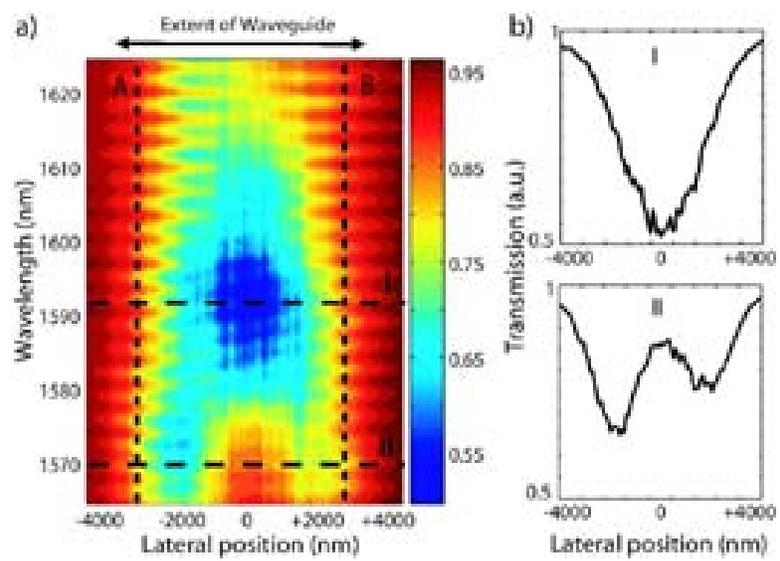

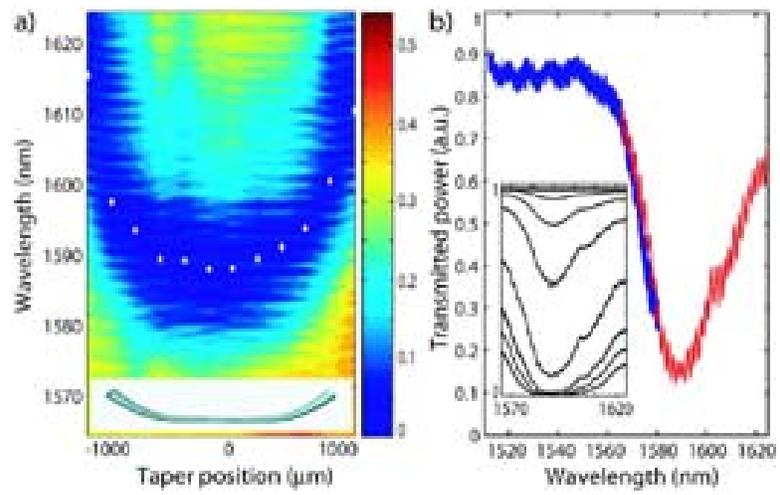